\documentclass[11pt]{article}

\usepackage[margin=1in]{geometry}

\usepackage[utf8]{inputenc} 
\usepackage[T1]{fontenc}    
\usepackage{lmodern}
\usepackage[ruled,vlined]{algorithm2e}
\usepackage{array}
\usepackage{caption}
\usepackage{enumitem}
\usepackage{subcaption}
\usepackage{lscape}         
\usepackage{hyperref}       
\usepackage{url}            
\usepackage{booktabs}       
\usepackage{nicefrac}       
\usepackage{microtype}      
\usepackage{multirow}
\usepackage[section]{placeins} 
\usepackage{xcolor}         
\usepackage{centernot}      
\usepackage{booktabs}
\usepackage{lipsum}
\usepackage{graphicx}
\usepackage{setspace}
\usepackage{amsmath,amsfonts,amssymb,amsthm,etoolbox,mathtools}
\usepackage{bm}             
\usepackage{accents}        
\usepackage{cleveref}       
\usepackage{longtable}      
\usepackage{authblk}
\usepackage{enumitem}
\usepackage{tikz}
\usepackage{verbatim}
\usepackage[all]{xy}

\DeclareMathOperator*{\argmin}{arg\,min}
\DeclareMathOperator*{\pr}{pr}

\newtheorem{theorem}{Theorem}[section]
\newtheorem*{theorem*}{Theorem}

\newtheorem{proposition}{Proposition}[section]
\newtheorem{lemma}{Lemma}[section]
\newtheorem{assumption}{Assumption}

\theoremstyle{definition}
\newtheorem{remark}{Remark}[section]
\newtheorem{example}{Example}[section]

\counterwithin{figure}{section}
\DeclareMathSizes{10}{10}{7}{4}
\allowdisplaybreaks
\usetikzlibrary{arrows.meta}
\usetikzlibrary{quotes,angles,positioning}
\usetikzlibrary{decorations.pathreplacing}
\usetikzlibrary{backgrounds}

\allowdisplaybreaks

\usepackage[
backend=biber,
style=authoryear,
citestyle=authoryear,
url=false,
uniquename=false,
]{biblatex}

\begin{document}


\title{Sample-constrained partial identification with application to selection bias}

\author[1,2,*]{Matthew J Tudball}
\author[1,2]{Rachael A Hughes}
\author[1,2]{Kate Tilling}
\author[3]{Jack Bowden}
\author[4]{Qingyuan Zhao}
\affil[1]{MRC Integrative Epidemiology Unit, University of Bristol}
\affil[2]{Population Health Sciences, Bristol Medical School, University of Bristol}
\affil[3]{College of Medicine and Health, University of Exeter}
\affil[4]{Statistical Laboratory, Department of Pure Mathematics and
  Mathematical Statistics, University of Cambridge}
\affil[*]{Correspondence to: \texttt{matt.tudball@bristol.ac.uk} \url{https://github.com/matt-tudball/selectioninterval}}

\maketitle

\begin{abstract}
Many partial identification problems can be characterized by the optimal value of a function over a set where both the function and set need to be estimated by empirical data. Despite some progress for convex problems, statistical inference in this general setting remains to be developed. To address this, we derive an asymptotically valid confidence interval for the optimal value through an appropriate relaxation of the estimated set. We then apply this general result to the problem of selection bias in population-based cohort studies. We show that existing sensitivity analyses, which are often conservative and difficult to implement, can be formulated in our framework and made significantly more informative via auxiliary information on the population. We conduct a simulation study to evaluate the finite sample performance of our inference procedure and conclude with a substantive motivating example on the causal effect of education on income in the highly-selected UK Biobank cohort. We demonstrate that our method can produce informative bounds using plausible population-level auxiliary constraints. We implement this method in the R package \texttt{selectioninterval}.
\end{abstract}

\section{Introduction} \label{sec:introduction}

\subsection{General problem} \label{sec:general-problem}

Partial identification problems arise when the observable data are only sufficient to identify a set or interval in which a parameter of interest is contained. A classical example from \textcite{Manski2003} is the missing data problem, where $Y$ is a discrete random variable and $S$ is a binary random variable indicating whether $Y$ is observed ($S = 1$) or not ($S = 0$). The distribution of $Y$ can be decomposed into 
\begin{equation} \label{eqn:canonical-example}
\pr(Y = y) = \pr(Y = y \mid S = 1) \pr(S = 1) + \pr(Y = y \mid S = 0) \pr(S = 0)
\end{equation}
for any $y$ in the support of $Y$. Given that $\pr(Y = y \mid S = 0)$ is unobserved, the smallest value that $\pr(Y = y)$ could take is $\pr(Y = y \mid S = 1) \pr(S = 1)$ and the largest value is $\pr(Y = y \mid S = 1) \pr(S = 1) + \pr(S = 0)$. Therefore, although $\pr(Y = y)$ itself cannot be point identified, it can be partially identified via the interval corresponding to the smallest and largest possible values.

As with the missing data example, many partial identification problems can be formulated as the optimal value of a population objective function, which we write as
\begin{equation}
    \nu = \inf \{Q(\theta) \colon \theta \in \Theta \},
\label{eqn:inference-target}
\end{equation} 
where $Q : \mathbb{R}^p \to \mathbb{R}$ and $\Theta \subseteq \mathbb{R}^p$. 

The field of stochastic optimization also considers problems of the form in \eqref{eqn:inference-target} and has built a large literature on estimation of, and inference to, $\nu$ when a sample analogue $Q_n$ is observed instead of $Q$ (where $n$ denotes the sample size). It will be convenient for us to frame the partial identification problem as a stochastic optimization problem and draw upon these existing results.

In this article we are primarily concerned with the more difficult setting where $\Theta$ must also be estimated empirically. We specifically consider a setting where $\Theta$ is characterized by inequality constraints of the form $\Theta = \{\theta \colon h_j(\theta) \leq 0, j = 1, \ldots, J \}$, where we may only observe corresponding estimators $h_{nj}(\theta)$. Within this setting, our goal is to find a lower confidence bound $C_n$ for any $0 < \alpha < 1$ such that 
\begin{equation}
    \lim_{n \to \infty} \pr(C_n \leq \nu) \geq 1 - \alpha,
\label{eqn:confidence-bound}
\end{equation}
which will suffice to provide useful statistical inference in a wide set of applications.

\subsection{Motivating application} \label{sec:motivating-application}

Our investigation is motivated by an applied question: how will selection bias affect the conclusions of population-based cohort studies? Many statistical analyses begin by selecting a study sample from some population of interest. When the sample is drawn non-randomly, then valid inference for the population is no longer guaranteed \parencite{Bareinboim2014}. Inverse probability weighting could be used to correct for this selection bias \parencite{Horvitz1952, Stuart2011}, but data on non-selected observations may be limited or unavailable altogether, such that the weights cannot be estimated. In such settings, there exist approaches to assess the sensitivity of estimates to a range of plausible inverse probability weights \parencite{Aronow2013, Thompson2014}. However, these approaches could be made more informative via a principled procedure for conducting statistical inference and the inclusion of relevant auxiliary information about the population. We demonstrate that such improvements can be made by casting these sensitivity analyses within the general framework described in \Cref{sec:general-problem}.

We are specifically motivated by studies conducted in UK Biobank, which is a large population-based cohort study widely analysed by health researchers. Studies of this cohort are potentially biased since recruited participants are known to differ systematically from the rest of the UK population on measures such as education, health status, age and geographical location \parencite{Fry2017, Hughes2019}.

\subsection{Existing literature} \label{sec:literature}

Statistical inference procedures have been developed for some special cases of our general problem in \cref{eqn:confidence-bound}. An area of particular focus is the so-called `sample average approximation' \parencite{Shapiro2009}. In this case, $Q$ is the expected value $Q(\theta) = E[f(\theta, X)]$ of some function $f$ and $Q_n$ is a sample average $Q_n(\theta) = n^{-1} \sum_{i=1}^n f(\theta, X_i)$, where $X$ is some random variable and $X_1, \ldots, X_n$ are independent draws of $X$. 

Statistical inference in the presence of $\Theta_n$ has been developed for convex sample average approximations, such that $f$ is convex in $\theta$ and $\Theta = \{ \theta \colon h_j(\theta) \leq 0, j = 1, \ldots, J \}$ where $h_j(\theta) = E[g_j(\theta, X)]$ and $g_j(\theta, X)$ is convex in $\theta$ for all $j$. \textcite{Shapiro1991} shows that the plug-in estimator
\begin{equation} \label{eqn:plug-in-estimator}
\nu_n^p = \inf \{ Q_n(\theta) \colon \theta \in \Theta_n \}.
\end{equation}
satisfies a central limit theorem under these convexity assumptions (and some additional regularity conditions), where $\Theta_n = \{ \theta \colon \sum_{i=1}^n g_j(\theta, X_i), j = 1, \ldots, J \}$. 

Moving away from convex problems, \textcite{Wang2008} consider the special case of minimizing a known function $Q$ subject to a single expected value constraint $\Theta =  \{ \theta \colon E[g(\theta, X)] \leq 0 \}$. Their approach consists of computing $\inf \{Q(\theta) \colon \theta \in \Theta_n \}$ and introducing $\Theta^{\epsilon} = \{ \theta \colon E[g(\theta, \xi)] \leq \epsilon \}$, where $\epsilon > 0$ represents a (small) deviation from the true problem. Sample size $n$ is then selected so that $\pr(\Theta^{-\epsilon} \subset \Theta_n \subset \Theta^{\epsilon})$ is sufficiently close to one.

This work also overlaps with the partial identification literature in econometrics, much of which considers inference for identified sets characterized by conditional or unconditional moment inequalities, commonly interpreted as the set of minimizers of some criterion function \parencite{Chernozhukov2007,Andrews2010,Andrews2013}. A related literature provides inference for parameters lying within partially identified sets, as opposed to inference for the set itself \parencite{Imbens2004, Stoye2009}. For a more comprehensive review of the partial identification literature, see \textcite{Molinari2020}.

\section{Confidence intervals for sample-constrained partial identification} \label{sec:ci-for-constrained-pi}
\subsection{Confidence intervals under known constraints} \label{sec:ci-under-pop-constraints}

In this section, we briefly summarize existing results on statistical inference for stochastic optimization when the set $\Theta$ is observed, which forms the basis of our generalization to situations where an estimate $\Theta_n$ of $\Theta$ is observed instead. Suppose the parameter space is defined by a set of inequality constraints
\begin{equation}
    \Theta = \bigl\{ \theta \colon h_j(\theta) \leq 0, \, j = 1, \ldots, J \bigr\}
\label{eqn:parameter-space}
\end{equation}
where an equality constraint for some $h_j(\theta)$ can be introduced by taking the inequality constraints of both $h_j(\theta)$ and $-h_j(\theta)$. Recall that our goal is to provide inference about the infimum $\nu = \inf \{Q(\theta) \colon \theta \in \Theta \}$.

Much of the literature in stochastic optimization is centered on the statistical properties of the estimator 
\begin{equation}
    \nu_n = \inf \{Q_n(\theta) \colon \theta \in \Theta \}.
    \label{eqn:sample-estimator}
\end{equation}

Consistency of optimal values and optimal solutions to such stochastic optimization problems is typically achieved by imposing uniform convergence of $Q_n(\theta)$ to $Q(\theta)$. First order asymptotic properties are obtained via the functional delta method. The key conditions are that the infimum, viewed as a function of $Q$, satisfies some notion of differentiability at $Q$ (see \textcite{Shapiro1991} for technical details) and that $n^{-1/2}(Q - Q_n)$ converges to a Gaussian process.

To make the previous discussion more concrete, consider the following four assumptions commonly placed on the stochastic optimization problem described above.
\begin{assumption} \label{assum:uniqueness}
The set of solutions to \eqref{eqn:inference-target} is a singleton $\{\theta \in \Theta \colon Q(\theta) = \nu \} = \{\vartheta\}$.
\end{assumption}

\begin{assumption} \label{assum:regularity}
Let $B \subseteq \mathbb{R}^p$ denote a compact set and $C(B)$ denote the space of continuous functions on domain $B$. Then  $\Theta \subseteq B$, $Q \in C(B)$ and $Q_n \in C(B)$ with probability one.
\end{assumption}

\begin{assumption} \label{assum:uniform-convergence}
$Q_n(\theta)$ converges to $Q(\theta)$ with probability one as $n \to \infty$ uniformly on $B$.
\end{assumption}

\begin{assumption} \label{assum:distribution-of-estimator}
As $n \to \infty$, the sequence $V_n(\theta) = n^{1/2}(Q(\theta) \, - \, Q_n(\theta))$ converges in distribution to a random element $V(\theta) \in C(B)$, where $V(\theta)$ is Gaussian process with mean 0 and variance $\sigma^2(\theta) \in C(B)$.
\end{assumption}

These assumptions are jointly sufficient to achieve consistency and asymptotic normality of $\nu_n$. 

\begin{proposition} \label{prop:consistency}
Let $\vartheta_n \in \argmin \{ Q_n(\theta) \colon \theta \in \Theta \}$ be a sample solution and $\nu_n$ be defined as in \cref{eqn:sample-estimator}. Under Assumptions \ref{assum:uniqueness}, \ref{assum:regularity} and \ref{assum:uniform-convergence}, $\nu_n \to \nu$ and $\vartheta_n \to \vartheta$ with probability one.
\end{proposition}

\Cref{prop:consistency} is identical to Theorem 5.3 in \textcite{Shapiro2009} under the condition that $\vartheta$ is unique.

\begin{proposition} \label{prop:asymptotic-normality}
Under Assumptions \ref{assum:uniqueness}, \ref{assum:regularity} and \ref{assum:distribution-of-estimator},
\begin{equation}
    n^{1/2} (\nu_n - \nu) \to \mathcal{N}\bigl(0, \sigma^2(\vartheta)\bigr)
    \label{eqn:asymptotic-normality} 
\end{equation}
in distribution, where $\sigma^2(\vartheta)$ is the asymptotic variance of $\nu_n$ defined in \Cref{assum:distribution-of-estimator}.
\end{proposition}

\Cref{prop:asymptotic-normality} is an immediate consequence of Theorem 3.2 in \textcite{Shapiro1991}. Although we do not restate the proof here, the intuition is that Assumptions \ref{assum:uniqueness} and \ref{assum:regularity} allow a notion of differentiability of the infimum and \Cref{assum:distribution-of-estimator} provides weak convergence of $n^{1/2} (Q - Q_n)$ to a Gaussian process, thus providing the conditions needed for an application of the delta method.

To use \Cref{prop:asymptotic-normality} to construct a valid confidence interval, we must take into consideration that both $\sigma^2$ and $\vartheta$ are unknown. To this end, we state an additional assumption followed by a proposition. 

\begin{assumption} \label{assum:variance}
There exists a uniformly strongly consistent estimator $\sigma_n^2(\theta) \in C(B)$ for $\sigma^2(\theta)$ such that $\sup_{\theta \in \Theta} |\sigma_n^2(\theta) - \sigma^2(\theta)| \to 0$ with probability one.
\end{assumption}

\begin{proposition} \label{prop:variance-consistency}
Under Assumptions \ref{assum:uniqueness}, \ref{assum:regularity}, \ref{assum:uniform-convergence} and \ref{assum:variance}, $\sigma^2_n(\vartheta_n) \, \to \, \sigma^2(\vartheta)$ with probability one.
\end{proposition}

The proof is located in \Cref{sec:appendix:proofs}. \Cref{assum:variance} applies uniform convergence to an estimator for the asymptotic variance of $Q_n(\theta)$. This strong notion of convergence for $\sigma_n^2(\vartheta_n)$ allows us to construct a confidence bound of the form
\begin{equation} \label{eqn:pop-confidence-bound}
    C_n = \nu_n - Z_{\alpha} \, \sigma_n(\vartheta_n) \, n^{-1/2}
\end{equation}
where $Z_{\alpha}$ is the upper $\alpha$-quantile of the standard normal distribution. This choice of $C_n$ has asymptotically exact nominal coverage $1-\alpha$ by \Cref{prop:asymptotic-normality}, \Cref{prop:variance-consistency} and Slutsky's theorem.

\subsection{Confidence intervals under sample constraints} \label{sec:ci-under-sample-constraints}

We now consider the more difficult setting where the constraint functions $h_j(\theta)$ need to be estimated as well. We instead observe an estimator $\Theta_n = \{ \theta \colon h_{nj}(\theta) \leq 0, j = 1, \ldots, J \}$ comprised of estimators of the constraint functions $h_{nj}(\theta)$. We will discuss what properties $\Theta_n$ must have to allow valid statistical inference for $\nu$.

It is tempting to follow the approach of the previous section and construct a plug-in estimator for $\nu$ by simply replacing $Q$ with $Q_n$ and $\Theta$ with $\Theta_n$ and finding the corresponding infimum (this is $\nu_n^p$ in \cref{eqn:plug-in-estimator}). The problem with this approach is that $\vartheta$ may lie inside $\Theta_n$ with very low probability, even asymptotically. This will prohibit the construction of a valid confidence interval for $\nu$ as illustrated by the contrived example below.

\begin{example} \label{ex:failure-of-naive}
Consider a problem of the form $Q(\theta) = \theta^2 + E[X]$ and $\Theta = \bigl\{\theta \colon \theta = E[X] \bigr\}$ where $X \sim \mathcal{N}(1,1)$ is a normally-distributed random variable. The plug-in estimators are $Q_n(\theta) = \theta^2 + \bar{X}_n$ and $\Theta_n = \{\theta \colon \theta = \bar{X}_n \}$, where $\bar{X}_n$ is the mean of $n$ independent and identically distributed draws of $X$. It follows that $\nu = 2$ and $\nu_n^p = \bar{X}_n^2 + \bar{X}_n $, where $\nu_n^p$ is the plug-in estimator \eqref{eqn:plug-in-estimator}. The asymptotic variance of $Q_n(\theta)$ is $\sigma^2(\theta) = 1$, which we assume is known. The resulting confidence bound is $C_n = \bar{X}_n^2 + \bar{X}_n - Z_{\alpha} n^{-1/2}$, where $C_n$ is the confidence bound \eqref{eqn:confidence-bound}. A simple Monte Carlo simulation demonstrates that the corresponding 95\% confidence interval for $n = 100$ exhibits sub-nominal coverage of around 70\%.
\end{example}

Existing approaches in stochastic optimization address the problem in \Cref{ex:failure-of-naive} by restricting to sample average approximations and imposing convexity of both $Q$ and $h$. To allow inference for a broader class of problems, we propose an intuitive but conservative approach which replaces $\Theta_n$ with an appropriate relaxation. In particular, we propose to use the relaxed set
\begin{equation} \label{eqn:relaxed-constraint}
    \Theta_n^r = \bigl\{ \theta \colon h_{nj}(\theta) \leq \epsilon_{nj}(\theta), \, j = 1, \ldots, J \bigr\}
\end{equation}
where $\epsilon_n(\theta) = (\epsilon_{n1}(\theta), \ldots, \epsilon_{nJ}(\theta))^T$ is some $J$-dimensional sequence such that $\epsilon_{nj}(\theta) \geq 0$ for all $\theta \in B$, chosen so that
\begin{equation} \label{eqn:constraint-coverage}
\lim_{n \to \infty} \pr(\Theta \subseteq \Theta_n^r) \geq 1 - \alpha_1
\end{equation} 
for some $0 < \alpha_1 < 1$. The exact forms of $\Theta_n^r$ and $\epsilon_n(\theta)$ are not crucial for our main results, provided \eqref{eqn:constraint-coverage} holds, which we discuss in more detail toward the end of this section.

Our proposed confidence bound is of the form $C_n(\theta) = Q_n(\theta) - Z_{\alpha_2} \sigma_n(\theta) n^{-1/2}$ for some $0 < \alpha_2 < 1$, where $Z_{\alpha_2}$ is the upper $\alpha_2$-quantile of the standard normal distribution.  We need to select a $\theta$ so that \cref{eqn:confidence-bound} is satisfied. This is accomplished by finding the optimal value and solution over the relaxed constraint set,
\begin{equation} \label{eqn:relaxed-optimum} \nu_n^r = \inf \{Q_n(\theta) \colon \theta \in \Theta_n^r \} \text{ and } \vartheta_n^r \in \argmin \{Q_n(\theta) \colon \theta \in \Theta_n^r \},
\end{equation}
and constructing a confidence bound of the form
\begin{equation} \label{eqn:relaxed-confidence-bound}
    C_n = C_n(\vartheta_n^r) = \nu_n^r - Z_{\alpha_2} \, \sigma_n(\vartheta_n^r) \, n^{-1/2}.
\end{equation}

We now need to demonstrate that $C_n$ covers $\nu$ with known probability in the limit. To this end, we need an additional technical assumption to hold.

\begin{assumption} \label{assum:variance-closeness}
Let $\zeta_n^r \in \argmin \{C_n(\theta) \colon \theta \in \Theta_n^r \}$ be the optimal solution of $C_n(\theta)$ over $\Theta_n^r$, then $\vert \zeta_n^r \, - \, \vartheta_n^r \vert$ converges to $0$ in probability.
\end{assumption}

\Cref{assum:variance-closeness} is imposed so that two important quantities become asymptotically close. The first quantity is $C_n(\zeta_n^r)$, which is the infimum over all confidence bounds in $\Theta_n^r$. This confidence bound is important because it provides a lower bound for other quantities with known coverage probabilities, which is a fact we utilize in our main result in \Cref{thm:coverage}. The second quantity is $C_n(\vartheta_n^r)$, which is our main confidence bound proposed in \eqref{eqn:relaxed-confidence-bound}.

We argue that \Cref{assum:variance-closeness} is reasonable in the sense that $\zeta_n^r$ and $\vartheta_n^r$ are solutions over two objective functions which converge uniformly to the same limit. To make this intuition more concrete, we provide some sufficient conditions for \Cref{assum:variance-closeness} in \Cref{sec:sufficient-conditions} and prove their sufficiency in \Cref{lemma:consistency}. Essentially, if \Cref{assum:uniform-convergence} is satisfied and $h_{nj}(\theta)$ and $\epsilon_{nj}(\theta)$ converge to $h_j(\theta)$ and 0 for all $j = 1, \ldots, J$ uniformly on $B$ with probability one, then we can show that both $\vartheta_n^r$ and $\zeta_n^r$ converge to $\vartheta$ with probability one. 

We claim that $C_n$ provides an asymptotically valid lower confidence bound.
\begin{theorem}
Suppose we select a relaxed constraint set $\Theta_n^r$ as in \cref{eqn:relaxed-constraint} and significance level $0 < \alpha_1 < 1$ such that $\lim_{n \to \infty} \pr(\Theta \subseteq \Theta_n^r) \geq 1 - \alpha_1$. Suppose we also select a significance level $0 < \alpha_2 < 1$. Then, under Assumptions \ref{assum:uniqueness} - \ref{assum:variance-closeness},
\begin{equation*}
\lim_{n \to \infty} \pr( C_n \leq \nu)  \, \geq \, 1 \, - \, \alpha_1 \, - \, \alpha_2.
\end{equation*} 
\label{thm:coverage}
\end{theorem}

The proof of \Cref{thm:coverage} can be found in \Cref{sec:appendix:proofs}. Here we outline the key steps in the proof. We begin by defining a deterministic sequence $\delta_n = Z_{\alpha_2} \epsilon n^{-1/2}$ where $\epsilon > 0$ is some small constant. We then show that $\pr( C_n \leq \nu)$ is bounded from below by the sum of two quantities: $\pr(C_n(\zeta_n^r) \leq \nu - \delta_n)$ and $\pr (\vert \sigma_n(\zeta_n^r) \, - \, \sigma_n(\vartheta_n^r) \vert \, \leq \, \epsilon ) - 1$. The second quantity  converges to $0$ under \Cref{assum:variance-closeness}. The remainder of the proof follows a similar argument to the main lemma of \textcite{Berger1994}. Whenever $\Theta \subseteq \Theta_n^r$, we know that $C_n(\zeta_n^r)$, which is the infimum over all confidence bounds in $\Theta_n^r$, will cover $\nu$ at least as often as $C_n(\vartheta_n)$, which is the confidence bound \eqref{eqn:pop-confidence-bound}. Therefore, $\pr(C_n(\zeta_n^r) \leq \nu, \Theta \subseteq \Theta_n^r) \geq \pr(C_n(\vartheta_n) \leq \nu, \Theta \subseteq \Theta_n^r)$. We also know that $\pr(C_n(\vartheta_n) \leq \nu, \Theta \subseteq \Theta_n^r) = \pr(C_n(\vartheta_n) \leq \nu) - \pr(C_n(\vartheta_n) \leq \nu, \Theta \not\subseteq \Theta_n^r)$ by the law of total probability. In the limit, the first probability on the right-hand side is equal to $1 - \alpha_2$ by \Cref{prop:asymptotic-normality} and the second probability is at most $\alpha_1$ by assumption. This allows us to arrive at our main result.

The proof sketch also provides some insight into why the naive plug-in estimator $\nu_n^p$ defined in \cref{eqn:plug-in-estimator} may fail to yield a valid confidence interval. A crucial quantity is $\pr(\Theta \subseteq \Theta_n^r)$, which is known under an appropriate choice of $\epsilon_n(\theta)$. The corresponding quantity for the plug-in estimator is $\pr(\Theta \subseteq \Theta_n)$, which could be arbitrarily small. In \Cref{ex:failure-of-naive}, this probability is zero.

It remains to discuss how to construct a relaxed set $\Theta_n^r$. Whenever $\Theta$ can be characterized by a set of moment inequalities (i.e. whenever $h_j(\theta) = E[m_j(\theta)]$), the moment inequalities literature summarized in \Cref{sec:literature} could be used to construct $\Theta_n^r$. A more conservative relaxed set could be constructed via an application of the intersection bound. Suppose the following assumption holds on the constraint functions:
\begin{assumption} \label{assum:constraint-convergence-distribution}
For all $\theta \in \Theta$ and $j = 1, \ldots, J$, $n^{1/2} \bigl(h_{nj}(\theta) \, - \, h_j(\theta) \bigr) \, \to \, \mathcal{N}\bigl(0, \, \sigma_j^2(\theta) \bigr)$ in distribution and $\sigma_{nj}^2(\theta)$ is a consistent estimator for $\sigma_j^2(\theta)$.
\end{assumption}
This fairly weak assumption means that $h_{nj}(\theta)$ is pointwise asymptotically normally distributed and that there is a consistent estimator for the variance. This assumption allows us to select
\begin{equation*}
\epsilon_n(\theta) = Z_{\alpha_{1j}} \, \sigma_{nj}(\theta) \, n^{-1/2}
\end{equation*}
where $\alpha_1 = \alpha_{11} + \alpha_{12} + \ldots + \alpha_{1J}$. It is straightforward to show that this choice of $\epsilon_n(\theta)$ satisfies \eqref{eqn:constraint-coverage}. We could shrink the size of $\Theta_n^r$ by assuming that $h_n(\theta) = (h_{1,n}(\theta), \ldots, h_{nj}(\theta))^{T}$ converges pointwise to a multivariate Gaussian with covariance matrix $\Sigma$ and consistent estimator $\Sigma_n$. This would allow us to construct $\Theta_n^r$ as an ellipsoid confidence region. 

\begin{remark} \label{remark:alpha-choice}
It remains to discuss how one would select $\alpha_1$ and $\alpha_2$. As a rule-of-thumb, we typically choose the midpoint $\alpha_1 = \alpha_2 = \alpha/2$. It is tempting to select $\alpha = \alpha_1 + \alpha_2$ and choose $C_n$ as the largest confidence bound over all $\alpha_1$ and $\alpha_2$ satisfying this equality. This would mean that $\alpha_1$ and $\alpha_2$ are sample-dependent quantities and so \Cref{thm:coverage} will not directly apply. However, we can reason heuristically that the best choice of $\alpha_1$ and $\alpha_2$ should lie at an interior point $\alpha_1 > 0$ and $\alpha_2 > 0$. For a fixed sample, as $\alpha_1 \to 0$ and $\alpha_2 \to \alpha$, $\Theta_n^r \to B$ and thus $C_n$ approaches the $100(1-\alpha)\%$ confidence interval over the unconstrained problem. As $\alpha_1 \to \alpha$ and $\alpha_2 \to 0$, $C_n \to -\infty$ and thus the confidence interval becomes arbitrarily wide.
\end{remark}

\begin{remark} \label{remark:two-sided}
So far we have focused on inference for the infimum, however, partial identification problems are often characterized by an identified set of the form $I = [\nu^l, \nu^u]$, where $\nu^l = \inf \{Q(\theta) \colon \theta \in \Theta \}$ and $\nu^u = \sup \{Q(\theta) \colon \theta \in \Theta \}$ \parencite{Imbens2004, Chernozhukov2013}. Suppose $\Theta_n^r$ is chosen so that $\pr(\Theta \subseteq \Theta_n^r) \geq 1 - \alpha_1/2$. Moreover, let $\nu_n^{r,l} = \inf\{Q_n(\theta) \colon \theta \in \Theta_n^r\}$ and $\nu_n^{r,u} = \sup\{Q_n(\theta) \colon \theta \in \Theta_n^r\}$ denote the optimal values and $\vartheta_n^{r,l}$ and $\vartheta_n^{r,u}$ denote the corresponding optimal solutions. The estimated interval can be written as $[\nu_n^{r,l}, \nu_n^{r,u}]$ and we can construct a confidence interval by combining the lower confidence bound for $\nu^l$ and the upper confidence bound for $\nu^u$, so that
\[
[\nu_n^{r,l} - Z_{\alpha_2/2} \sigma_n(\vartheta_n^{r,l}) n^{-1/2}, \nu_n^{r,u} + Z_{\alpha_2/2} \sigma_n(\vartheta_n^{r,u}) n^{-1/2}]
\]
will cover $I$ with probability at least $1-\alpha_1-\alpha_2$. This is the two-sided analogue of the one-sided confidence interval proposed in \cref{eqn:relaxed-confidence-bound} and \Cref{thm:coverage}.
\end{remark}

\section{Sensitivity analysis via a logistic model} \label{sec:logistic}

\subsection{Set-up} \label{sec:motivating-application:set-up}

We now return to the motivating example of selection bias in population-based cohort studies briefly described in \Cref{sec:motivating-application}. Specifically, we generalize the sensitivity analysis proposed in \textcite{Thompson2014}, who define a logistic model for the probability of sample selection and propose to select parameters based on domain knowledge, or enumerate a large number of possible parameters. This approach is challenging to implement in the presence of complicated selection mechanisms with many parameters. Plausible sets of parameters that introduce bias in estimates of interest may be overlooked. Therefore, we begin by framing \textcite{Thompson2014} as an optimization problem over a space of plausible parameters, and describe how relevant auxiliary information could be introduced to further restrict the parameter space and provide more informative bounds (e.g. survey response rates, population means and negative controls). An additional sensitivity analysis, \textcite{Aronow2013}, is generalized in \Cref{sec:al-extension}.

Consider an independent and identically distributed draw of size $N$ from an infinite population. For concreteness, we can think of this finite draw as the set of individuals who are eligible to enter the sample. Let $S_i \in \{0, 1\}$ be a selection indicator for whether individual $i$ enrols in the sample, where $S_i = 1$ indicates sample participation, and let the observed sample size be denoted by $n = \sum_{i=1}^N S_i$. For notational convenience, we assume $S_1 = \ldots = S_n = 1$ and $S_{n+1} = \ldots = S_N = 0$. 

Within the observed sample (i.e. for $S_i = 1$), we observe a vector of variables related to sample selection $W_i \in \mathbb{R}^K$. As in \textcite{Thompson2014}, we assume that the probability of sample selection admits a logistic form,
\begin{equation} \label{eqn:logit}
e(W_i) = \pr(S_i = 1 \mid W_i) = \frac{\exp(\theta_0 + \theta_1 W_{i1} + \ldots + \theta_k W_{iK})}{1 + \exp(\theta_0 + \theta_1 W_{i1} + \ldots + \theta_k W_{iK})},
\end{equation}
where we define $\theta = (\theta_0, \theta_1, \theta_2 \ldots, \theta_K)^T$. 

For illustration, suppose that our parameters and estimators of interest are (respectively) characterized by the solution to the instrumental variable moment conditions
\begin{equation} \label{eqn:moment-conditions}
0 = E[Z_i(Y_i - \beta(\theta)^T X_i)/e(W_i) \mid S_i = 1], \quad 0 = \sum_{i=1}^n Z_i (Y_i - \beta_n(\theta)^T X_i) / e(W_i),
\end{equation}
where $Z_i \in \mathbb{R}^L$ is a vector of instrumental variables, $X_i \in \mathbb{R}^L$ is a vector of explanatory variables and $Y_i \in \mathbb{R}$ is a response variable \parencite{Hayashi2000}. This is equivalent to the least squares moment conditions whenever $Z_i = X_i$. The moment conditions are inverse weighted by the probability of sample selection $e(W_i)$; this will weight the sample back to the population implied by the weights under a missing-at-random assumption that $(Z_i,X_i,Y_i) \perp S_i \mid W_i$. For this reason, we write the parameters $\beta(\theta)$ and estimators $\beta_n(\theta)$ as functions of the parameters $\theta$ that characterize the sample selection probabilities. We also allow the instruments, explanatory variables and response (and their higher order and interaction terms) to be elements of $W_i$.

Without loss of generality, suppose that our parameter of interest is the first element $\beta_1(\theta)$ since we consider optimization of a scalar-valued function in our framework.

\subsection{Sensitivity parameters} \label{sec:sensitivity-parameters}

The key challenge is that we only observe those for whom $S_i = 1$ so that the parameters $\theta$ cannot be estimated. A sensible approach is to place bounds on the possible values of $\theta$ and identify the smallest and largest values that the corresponding estimator $\beta_{1n}(\theta)$ could take. Since we have assumed a logistic form for the selection probabilities \eqref{eqn:logit}, we can select sensitivity parameters which have a natural interpretation in terms of odds ratios. 

Without loss of generality, suppose each $W_{ik}$ has mean zero and standard deviation one within the sample. We can then choose a parameter $\Lambda_1 \geq 1$ such that
\begin{equation} \label{eqn:lambda1}
    \Lambda_1^{-1} \leq \exp(\theta_j) \leq \Lambda_1, ~ k = 1, \ldots, K.
\end{equation}
We can interpret $\Lambda_1$ as the change in the conditional odds of sample selection from a one standard deviation increase in $W_{ik}$, holding all else fixed. When $\Lambda_1 = 1$, sample selection is completely random. Of course, we could select sensitivity parameters $\Lambda_{1k}$ on a variable-by-variable basis for $k = 1, \ldots, K$, although choosing a single $\Lambda_1 = \max_k \Lambda_{1k}$ simplifies the interpretation of the sensitivity analysis. 

The intercept term $\theta_0$ also needs to be bounded. We can choose two parameters $\Lambda_0^l, \Lambda_0^u \in (0, 1)$ such that
\begin{equation} \label{eqn:lambda0}
    \Lambda_0^l \leq \exp(\theta_0) \leq \Lambda_0^u
\end{equation}
which can be interpreted as the odds of sample selection among those for whom $W_{ik} = 0$ for all $k$.

Rearranging \cref{eqn:lambda1,eqn:lambda0} shows that the sensitivity parameters $(\Lambda_0^l, \Lambda_0^u, \Lambda_1)$ characterize a compact subset of $\mathbb{R}^{K+1}$,
\begin{equation} \label{eqn:example-theta}
\theta \in \Theta = [\log(\Lambda_0^l), \log(\Lambda_0^u)] \times [\log(1/\Lambda_1), \log(\Lambda_1)]^K
\end{equation}

From here, we can define the estimand and estimator (respectively) for the worst-case lower bound of $\beta_1(\theta)$ as
\[
\nu = \inf \{ \beta_1(\theta) \colon \theta \in \Theta \}, \quad \nu_n = \inf \{ \beta_{1n}(\theta) \colon \theta \in \Theta \}.
\]
We could of course estimate the worst-case upper bound for $\beta_1(\theta)$ by taking the supremum of $\beta_{1n}(\theta)$ over $\Theta$ (see \Cref{remark:two-sided}). 

\subsection{Auxiliary information constraints} \label{sec:auxiliary}

We now introduce several common examples where there may be discordance between known population quantities and quantities implied by the inverse probability weights. In general, provided we can formulate the constraints as a statistical test with a known asymptotic distribution, they can be placed within our framework. 
\begin{example} \label{ex:response-rate}
Suppose we know the response rate for a survey-based sample $r = E[e(W_i)]$. It is straightforward to show that $E[1/e(W_i) \mid S_i = 1] = 1/r$. This means that the within-sample expectation of the inverse selection probabilities is equal to the inverse response rate. The response rate constraint can be formulated as
\begin{equation} \label{eqn:response-rate-constraint}
h_{nj}(\theta) = \frac{1}{n} \sum_{i=1}^n (1/e(W_i) - 1/r) \leq Z_{\alpha_{1j}/2} \, \sigma_{nj}(\theta) / n^{1/2},
\end{equation}
where $\sigma_{nj}(\theta)$ is the sample variance of $1/e(W_i)$.
\end{example}

\begin{example} \label{ex:covariate-mean}
Suppose we know the population mean $E[W_{ik}]$ of some $W_{ik} \in W_i$. The inverse probability weighted sample mean of $W_{ik}$ should therefore equal this mean in expectation, since
\[
\frac{E[W_{ik}/e(W_i) \mid S_i = 1]}{E[1/e(W_i) \mid S_i = 1]} = E[W_{ik}].
\]
This is conceptually similar to the raking procedure in survey sampling \parencite{EdwardsDeming1940}, which adjusts sampling weights to match known marginal totals. The covariate mean constraint can be formulated as
\begin{equation} \label{eqn:covariate-mean-constraint}
h_{nj}(\theta) = \frac{1}{n} \sum_{i=1}^n (W_{ik} - E[W_{ik}])/e(W_i) \leq Z_{\alpha_{1j}/2} \, \sigma_{nj}(\theta) / n^{1/2},
\end{equation}
where $\sigma_{nj}(\theta)$ is the sample variance of $(W_{ik}-E[W_{ik}])/e(W_i)$. 
\end{example}

\begin{example} \label{ex:direction}
Suppose we are confident that higher values of $W_{ik}$ are associated with an increased probability of sample selection. For example, $W_{ik}$ could be years of education and we might know from comparisons with representative samples (e.g. census) that better educated individuals are more likely to select into our sample, conditional on other selection variables, so that $\theta_j \geq 0$ \textit{a priori}. 
\end{example}

\begin{example} \label{ex:negative-control}
Suppose we know that two variables $W_{ik}$ and $W_{ik^{\prime}}$ are uncorrelated in the population. The inverse probability weighted correlation between $W_{ik}$ and $W_{ik^{\prime}}$ should therefore be zero. For example, due to the independent assortment of chromosomes, biological sex and autosomal genetic variants should be independent in the population, however, \textcite{Pirastu2021} demonstrate that there is significant correlation within UK Biobank. This constraint can be formulated in several ways, for example fixing the regression coefficient of $W_{ik}$ on $W_{ik^{\prime}}$ to be zero.
\end{example}

\Cref{ex:response-rate,ex:covariate-mean,ex:negative-control} are two-sided constraints such that we also want these inequalities to hold for $-h_{nj}(\theta)$.

\begin{remark}\label{remark:shape-constraints}
In the population means setting, \textcite{Miratrix2018} demonstrate how to place shape constraints on the weighted empirical distribution of the response. Their approach involves constructing the worst-case weighted distribution given the \textcite{Aronow2013} bounding assumptions (see \Cref{sec:al-extension}). This results in a set which contains the oracle weighted distribution with probability approaching one. 

Provided we have a valid test, we can implement shape constraints within our framework without the need to characterize the worst-case weighted distribution. In the simplest case, we might want a variable to follow a known distribution in the population. For example, the distribution of IQ scores should be normal with mean 100 and standard deviation 15, which is a stronger constraint than \Cref{ex:covariate-mean}. This could be formulated as a Kolmogorov-Smirnov test and the relaxed constraint set could be constructed via the null distribution of that test. 
\end{remark}

\section{Simulations} \label{sec:simulation}

The aim of these simulations is to provide a brief assessment of the finite sample and limiting properties of the inference procedure described in \Cref{sec:ci-for-constrained-pi}. For concreteness, we simulate the sensitivity analysis for selection bias described in \Cref{sec:logistic}. Our parameter $\beta_1(\theta)$ and estimator $\beta_{n1}(\theta)$ are the coefficient of a linear regression, specifically the solution to the least squares moment conditions in \cref{eqn:moment-conditions} for $(X_i, Y_i) \sim \mathcal{N}(0,I_2)$, where $I_2$ is the identity matrix, and $Z_i = X_i$. The selection variables are $W_i = (X_i, Y_i)$. 

We consider three distinct scenarios for the constraints. In the first scenario, we impose only sensitivity parameters $\Lambda_0^l = 0.11$, $\Lambda_0^u = 0.25$ and $\Lambda_1 = 3$. In the second scenario, we also impose a direction constraint $\theta_1 \geq 0$ via \Cref{ex:direction}. In the third scenario, we impose both the previous direction constraint and set the response rate equal to $0.15$ via \Cref{ex:response-rate}. In each scenario, we use the discussion in \Cref{remark:two-sided} to construct a two-sided 95\% confidence interval for the identified set $I = [\nu^l, \nu^u]$, where $\nu^l = \inf\{\beta_1(\theta) \colon \theta \in \Theta\}$, $\nu^u = \sup\{\beta_1(\theta) \colon \theta \in \Theta\}$ and $\Theta$ is of the form in \cref{eqn:example-theta}. The first and second scenarios have no sample constraints and so the confidence interval corresponds to the one in \cref{eqn:pop-confidence-bound}. The third scenario employs the confidence interval proposed in \cref{eqn:relaxed-confidence-bound} and \Cref{thm:coverage}. 

Each scenario has distinct properties. In the first scenario, there are two solutions to the population optimization problems, thus violating \Cref{assum:uniqueness}. In the second scenario, the addition of a direction constraint rules out one of the two solutions and satisfies \Cref{assum:uniqueness}. In the third scenario, the introduction of a sample constraint necessitates the use of our relaxed confidence bound. In this scenario, we use our rule-of-thumb from \Cref{remark:alpha-choice} to select $\alpha_1 = \alpha_2 = 0.025$ for both the upper and lower bounds of the two-sided confidence interval.

\Cref{tab:simulation-results} summarizes the results and broadly aligns with our theoretical predictions. The first scenario violates \Cref{assum:uniqueness} and the impact of this violation is substantial over-coverage of the confidence interval. Intuitively, this occurs because the sample solution $\vartheta_n$ will occur at (or near) the population solution that happens to minimize $Q_n(\theta)$ in that particular sample, which will result in a systematically wider confidence interval. The second scenario satisfies all assumptions for \Cref{prop:asymptotic-normality} and therefore converges to exact nominal coverage. The third scenario imposes a sample constraint and exhibits some over-coverage. This over-coverage can occur because our confidence bound in \Cref{thm:coverage} sidesteps the covariance between the constraints $h_{nj}(\theta)$ and objective function $Q_n(\theta)$, instead imposing a worst-case intersection bound. 

\begin{table}[!ht]
   \begin{center}
     \begin{tabular}{c|ccccccc}
       \toprule
       & \multicolumn{7}{c}{Sample size} \\
       Scenario & 10 & 25 & 50 & 100 & 200 & 500 & 1000 \\
       \midrule
       1 & 0.972 & 0.992 & 0.995 & 0.997 & 0.998 & 0.996 & 0.995  \\
       2 & 0.936 & 0.974 & 0.981 & 0.983 & 0.979 & 0.966 & 0.947 \\
       3 & 0.953 & 0.985 & 0.991 & 0.991 & 0.987 & 0.986 & 0.979 \\
       \bottomrule
     \end{tabular}
     \caption{Coverage frequency for the three scenarios over 5000 Monte Carlo replications.}
     \label{tab:simulation-results}
   \end{center}
 \end{table}

\section{Applied example: effect of education on income} \label{sec:education-on-income}

We consider an instrumental variable design looking at the effect of education on income in the UK Biobank cohort. Our instrument is exposure to an educational reform taking place in England in 1972. Our exposure is whether an individual remained in school at least until age 16 and our outcome is whether an individual earned more than \pounds 31,000 per year in 2006. We restrict our sample to individuals who turned 15 within 12 months of September 1972 and we control for sex and month-of-birth indicators. The unweighted estimate is 0.18 (95\% confidence interval 0.08 - 0.28). An in-depth exposition of this design can be found in \textcite{Davies2018} and \Cref{sec:applied-example}. 

To address this, we first apply the sensitivity analysis described in \Cref{sec:sensitivity-parameters} without auxiliary constraints, where the probability weights contain sex, years of education, income, age, days of physical activity per week and an interaction term between education and income. We choose sensitivity parameters $\Lambda_0^l = 0.02$, $\Lambda_0^u = 0.25$ and $\Lambda_1 = 2$, so that the average individual in the sample has an odds of sample selection between $0.02$ and $0.25$ and each variable in the model can induce a marginal odds of sample selection between $0.5$ and $2$. Using these parameters, our sensitivity analysis suggests that the effect estimate lies in the interval $[-1.34, 0.94]$ (95\% confidence interval $-1.84, 1.29]$). This interval is completely uninformative as it spans the full range of possible estimates.

One explanation for this conservativeness is that this simple sensitivity analysis does not utilize all of the information available to us on the target population and the sample selection mechanism. The minimizing (maximizing) weights corresponding to this interval imply that the proportion of males in the population is 38.52\% (46.6\%) and the proportion of households with a gross income greater than \pounds 31000 is 95.84\% (95.66\%), all of which are inconsistent with known characteristics of the UK population.

We consider four constraints that are typical of the information available to applied researchers using datasets such as UK Biobank. The first constraint is the response rate of UK Biobank (5.5\%), which is the proportion of individuals who entered the cohort after receiving an invitation. The second constraint is the proportion of males in the UK population within the UK Biobank age range of 40-69 (49.5\%). The third is the proportion of UK households earning more than \pounds 31000 per year at the date of UK Biobank recruitment in 2006 (21\%). The fourth is the average age of individuals within our 2 year age bracket (48.98). All statistics were obtained from publicly available records from the UK's Office of National Statistics.

\begin{figure}[ht] 
  \begin{center}
    \includegraphics[scale=0.65]{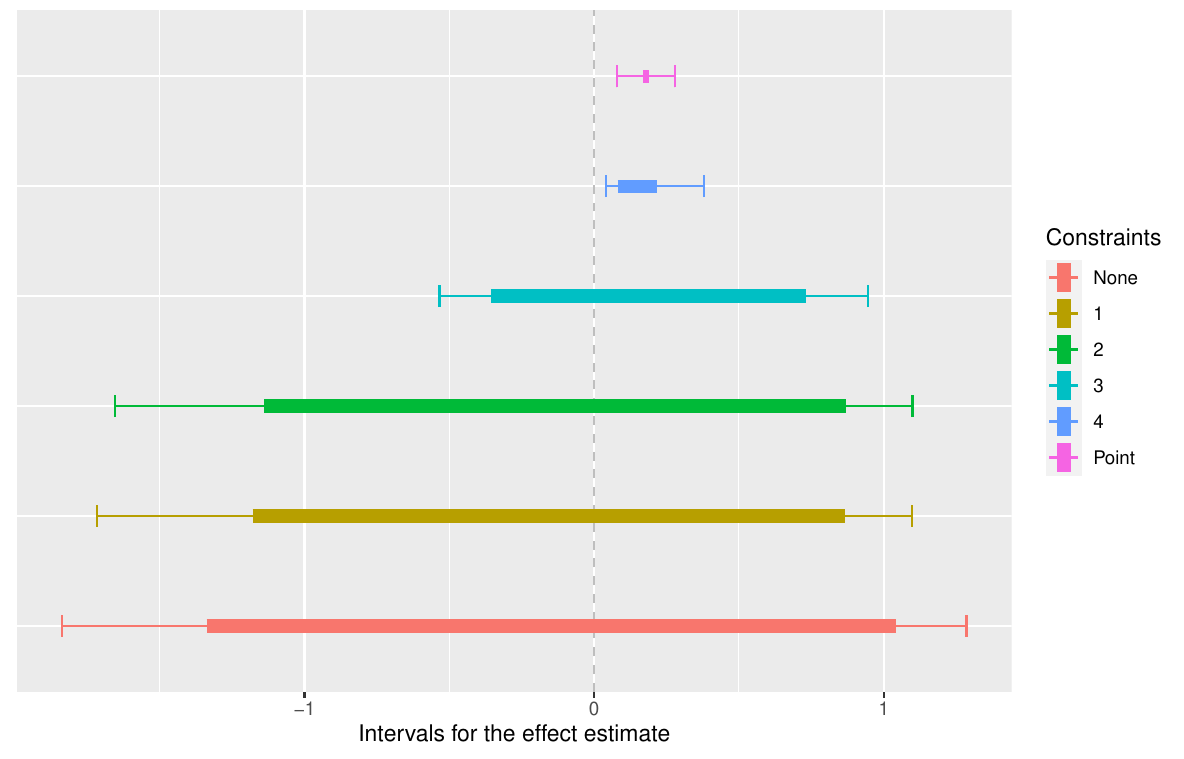}
     \caption{Estimated intervals (thick lines) and corresponding confidence intervals (thin lines) for effect estimates in the applied example. Point represents the unweighted point estimate. Each constraint is added sequentially. No constraint means that only the sensitivity parameters $\Lambda_0^l = 0.02$, $\Lambda_0^u = 0.25$ and $\Lambda_1=2$ are imposed. Constraint 1 sets the response rate equal to 5.5\%. Constraint 2 sets the proportion of males in the population to be 49.5\%. Constraint 3 sets the proportion of households earning more than \pounds 31000 to be 21\%. Constraint 4 sets the average age of individuals to be 48.98 years.} \label{fig:applied-example}
  \end{center}
\end{figure}

\Cref{fig:applied-example} shows the resulting estimated intervals (thicker lines) and their corresponding confidence intervals (thinner lines) where each constraint is added sequentially. The estimated intervals and confidence intervals correspond to those described in \Cref{remark:two-sided}. We can see that each additional constraint reduces the width of the interval, with constraints 3 and 4 (the household income and age constraints respectively) seemingly having the largest marginal impact. The top interval includes all constraints and is quite informative for the desired effect, rejecting the null and suggesting an effect estimate in the range 0.08 - 0.22 (95\% confidence interval 0.04 - 0.38). The unweighted estimate of 0.18 (95\% confidence interval 0.08 - 0.28) reported in \Cref{sec:applied-example} still lies within this interval, but our sensitivity analysis suggests some increased uncertainty in the range of effect estimates. These results also suggest that, despite the potential conservativeness of the confidence interval in \Cref{thm:coverage}, it can still produce informative bounds in practice.

\section{Discussion} \label{sec:discussion}

There has been some existing work on bootstrap inference for Rosenbaum-type sensitivity analyses \parencite{Zhao2019}. This approach considers a fixed parameter space $\Theta$. It is unclear how to select the relaxation parameter $\epsilon_n(\theta)$ in a bootstrap analogue of our method under estimated constraints. Simple approaches, such as constructing $\Theta_n^r$ via asymptotic approximations and then bootstrapping the distribution of $\nu_n^r$, are plausible but their statistical properties remain to be explored. 

In some instances, the target of inference is $Q(\theta^*)$, where $\theta^*$ is some true parameter lying within $\Theta$, rather than $\nu$. Suppose $\Theta$ is known and we have a two-sided identified set $[\inf_{\theta \in \Theta} Q(\theta), \, \sup_{\theta \in \Theta} Q(\theta)]$, then if $Q(\theta^*)$ lies near the boundary of this set (and the set has positive width), the non-coverage probability of the corresponding confidence interval is effectively one-sided in the limit. A naive two-sided confidence interval constructed around the identified set may be too conservative. \textcite{Imbens2004} discuss approaches for maintaining uniform coverage of $Q(\theta^*)$. The central limit theorem established by \textcite{Shapiro1991} for known $\Theta$ is amenable to their framework (although, to our knowledge, has not been formally used in this setting); extending this result to sample-constrained problems would be a valuable contribution. \textcite{Stoye2009} further extends \textcite{Imbens2004} by developing confidence intervals that exhibit uniform coverage for $Q(\theta^*)$ without relying on assumed superefficiency of the estimated interval width. 

\printbibliography

\section*{Acknowledgements}
The authors thank the referees and associate editor for their helpful and detailed comments. This research has been conducted using the UK Biobank Resource. UK Biobank received ethical approval from the Research Ethics Committee (REC reference for UK Biobank is 11/NW/0382). This research was approved as part of application 8786. Tudball would like to acknowledge financial support from the Wellcome Trust (grant number 220067/Z/20/Z). Bowden is supported by an Expanding Excellence in England (E3) award from the University of Exeter.  For the purpose of Open Access, the author has applied a CC BY public copyright licence to any Author Accepted Manuscript version arising from this submission.

\newpage

\appendix

\section{Further details for the applied example} \label{sec:applied-example}
\subsection{Description of the design} \label{sec:design-description}
We implement an instrumental variable design for the effect of education on income in the UK Biobank cohort \parencite{Davies2018}. The proposed instrument is based on a September 1972 education reform in England which raised the school leaving age from 15 to 16. Individuals who turned 15 just prior to the implementation of this reform were allowed to leave school, while individuals who turned 15 just after were required to remain in school until they were 16. This created a sharp discontinuity in the policies that the two groups were exposed to. Under the assumption that individuals on either side of the age threshold are otherwise identical, we can use this policy reform as an instrumental variable.

\subsection{Varying the sensitivity parameters} \label{sec:varying-parameters}
It is important to report a few choices for the sensitivity parameters to understand which parameters are driving the width of the interval. Selecting $\Lambda_1 < 1.75$ results in an empty constraint set, indicating that there are no parameters which satisfy all of the auxiliary information constraints provided. The response rate constraint of \Cref{ex:response-rate} appears to be more informative when the interval $(\Lambda_0^l, \Lambda_0^u)$ is wider, which is expected. For all choices of parameters we consider, the constraints are informative and recover an interval that rejects the null.

\begin{figure}
  \begin{subfigure}[t]{0.48\textwidth}
\includegraphics[scale=0.4]{applied_example_1.pdf}
\caption{$\Lambda_0^l = 0.02, \Lambda_0^u = 0.2, \Lambda_1 = 2$}
\label{fig:sim1}
  \end{subfigure} \hfill
  \begin{subfigure}[t]{0.48\textwidth}
 \includegraphics[scale=0.4]{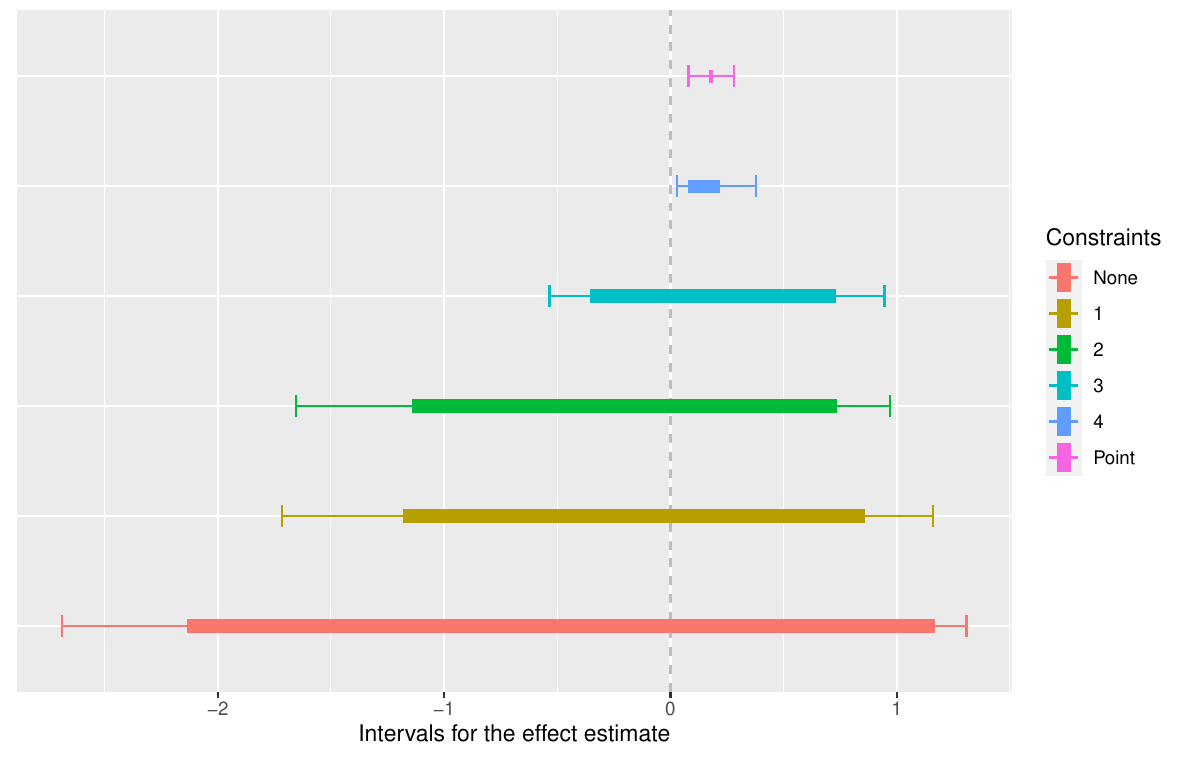}
    \caption{$\Lambda_0^l = 0.01, \Lambda_0^u = 0.5, \Lambda_1 = 2$}
    \vspace{1ex}
    \label{fig:sim2}
  \end{subfigure}\\[0.5em]
 
  \begin{subfigure}[t]{0.48\textwidth}
    \includegraphics[scale=0.4]{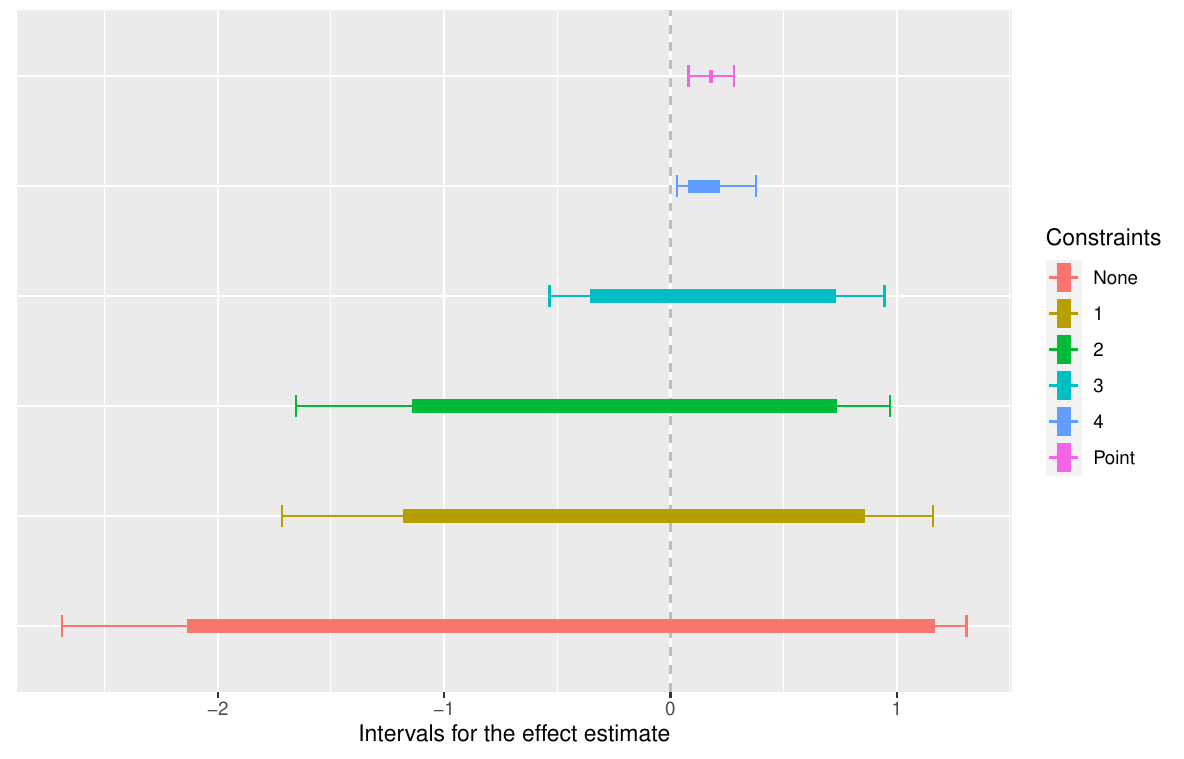}
    \caption{$\Lambda_0^l = 0.02, \Lambda_0^u = 0.2, \Lambda_1 = 2.5$}
    \label{fig:sim3}
  \end{subfigure} \hfill
  \begin{subfigure}[t]{0.48\textwidth}
    \includegraphics[scale=0.4]{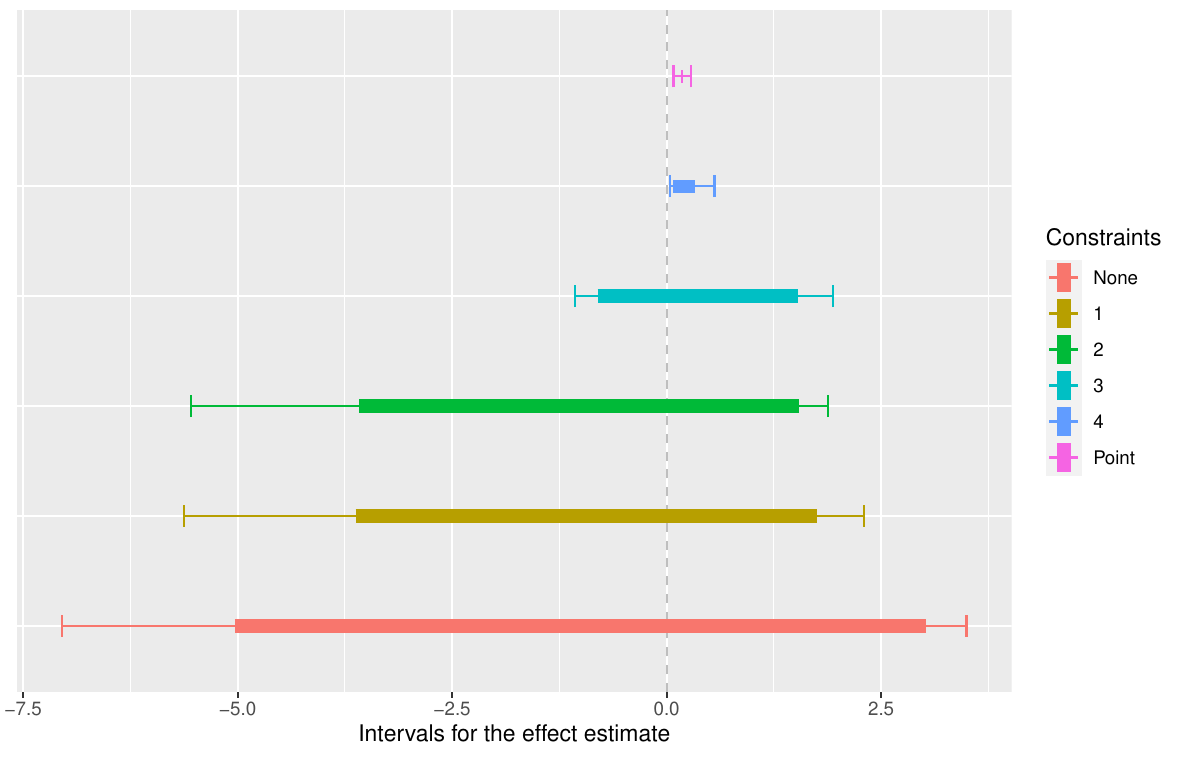}
    \caption{$\Lambda_0^l = 0.01, \Lambda_0^u = 0.5, \Lambda_1 = 2.5$}
    \label{fig:sim4}
  \end{subfigure}\\[0.5em]
  
  \begin{subfigure}[t]{0.48\textwidth}
    \includegraphics[scale=0.4]{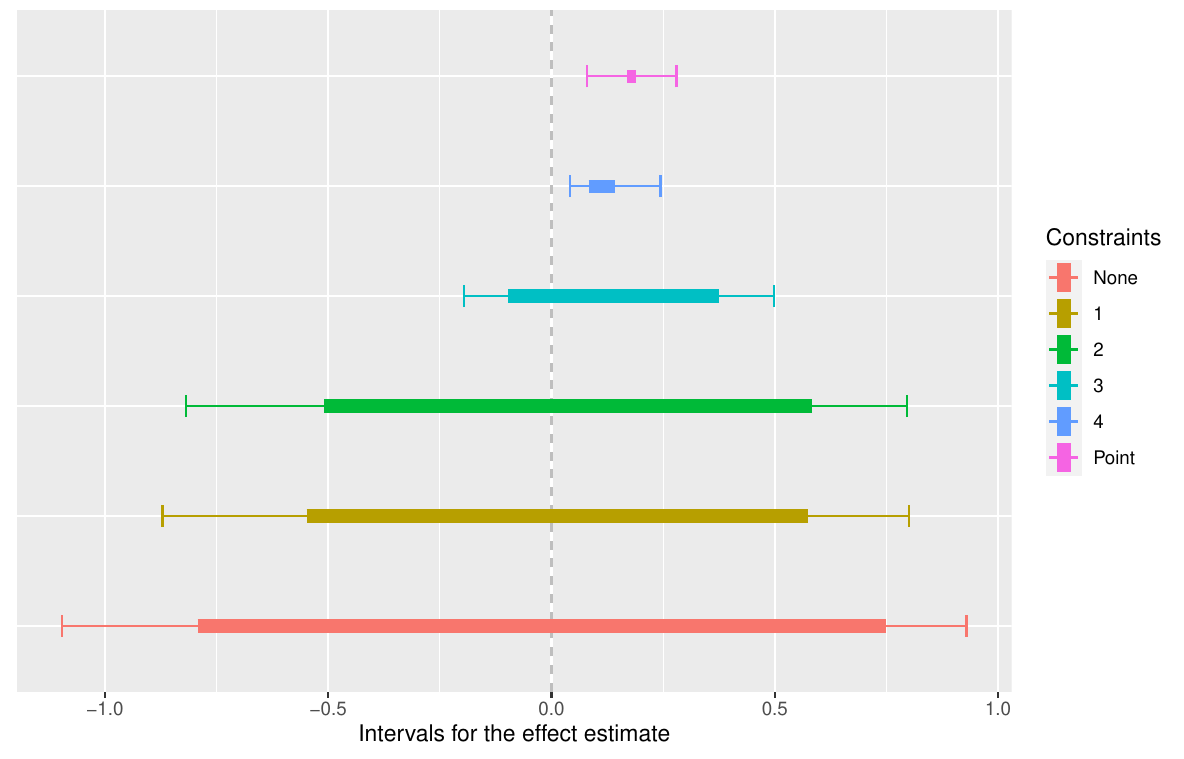}
    \caption{$\Lambda_0^l = 0.02, \Lambda_0^u = 0.2, \Lambda_1 = 1.75$}
    \label{fig:sim5}
  \end{subfigure} \hfill
  \begin{subfigure}[t]{0.48\textwidth}
    \includegraphics[scale=0.4]{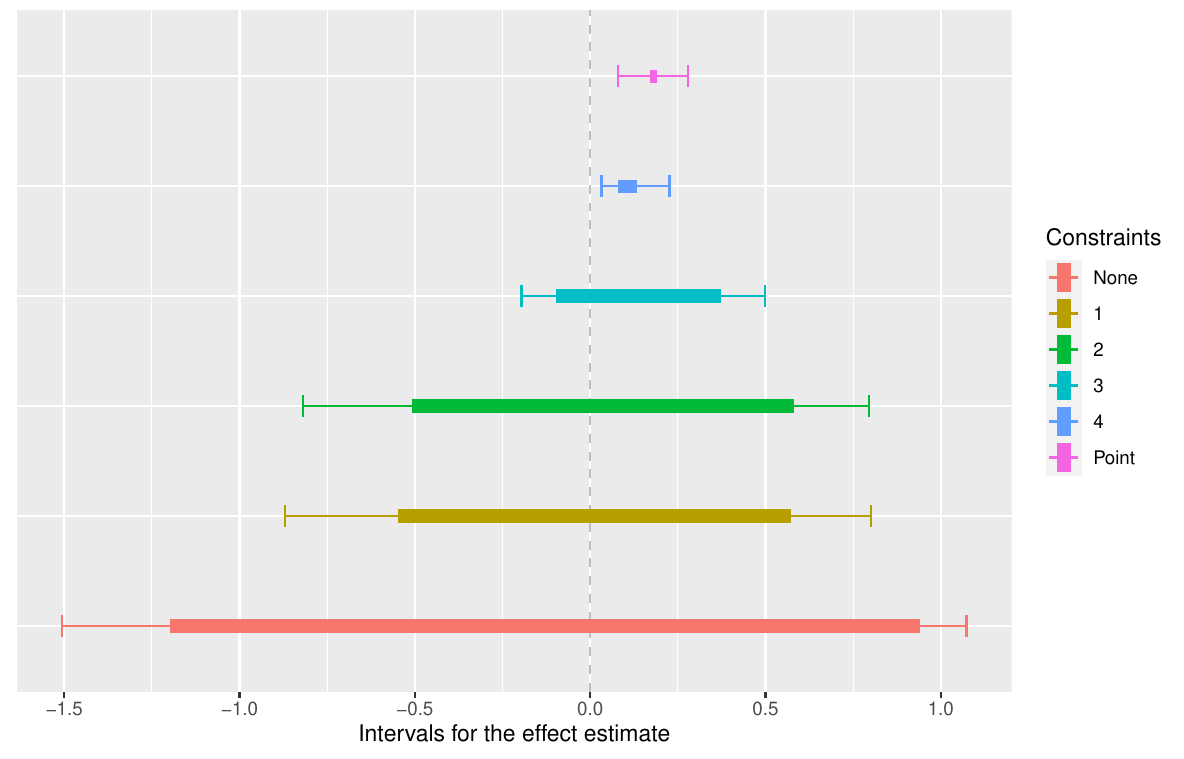}
    \caption{$\Lambda_0^l = 0.01, \Lambda_0^u = 0.5, \Lambda_1 = 1.75$}
    \label{fig:sim6}
  \end{subfigure}
  \caption{This figure presents several choices of sensitivity parameters for the applied example described in \Cref{sec:education-on-income}.}\label{fig:varying-parameters}
\end{figure}

\subsection{Visualizing the feasible region} \label{sec:constraint-visual}

It is also illustrative to plot the feasible region for a couple of simple examples. Suppose that $W_i$ consists only of the sex variable. We select sensitivity parameters $(\Lambda_0^l, \Lambda_0^u, \Lambda_1) = (0.02, 0.2, 2)$ as usual and consider two constraints: setting the response rate to be 0.055 and setting the population mean of male sex to be 0.495.    

\begin{figure}
  \begin{subfigure}[t]{0.48\textwidth}
    \includegraphics[scale=0.75]{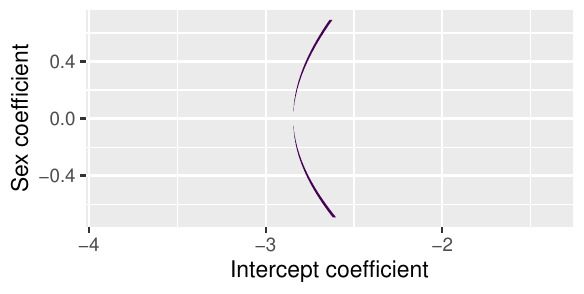}
    \caption{Response rate constraint}
    \label{fig:constraint1}
  \end{subfigure} \hfill
  \begin{subfigure}[t]{0.48\textwidth}
    \includegraphics[scale=0.75]{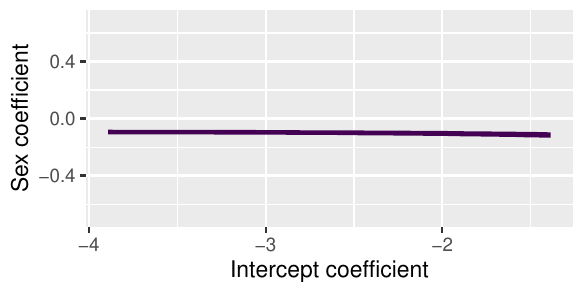}
    \caption{Covariate mean constraint}
    \label{fig:constraint2}
  \end{subfigure}
  \caption{This figure plots the feasible region (purple region) for a simple selection model with one variable and an intercept. Panel (a) sets the response rate to be 0.055 and panel (b) sets the population mean of male sex to be 0.495.} \label{fig:constraint-set}
\end{figure}

\Cref{fig:constraint-set} plots the two feasible regions. The feasible regions are both small in comparison to the space implied by the sensitivity parameters. Imposing both constraints simultaneously will result in a non-empty feasible region. In fact, \textcite{Nevo2003} shows that the parameters of the selection model are exactly identified in this case provided each constraint provides a unique restriction on $\theta$, in the sense that the outer product of the corresponding equality constraints is of full rank. 

\section{Sufficient conditions for \texorpdfstring{\Cref{assum:variance-closeness}}{Assumption 5}} \label{sec:sufficient-conditions}

\begin{assumption} \label{assum:compact-subset}
For sufficiently large $n$, $\Theta_n^r \subseteq B$ with probability one.
\end{assumption}

\begin{assumption} \label{assum:constraint-convergence}
For all $j = 1, \ldots, J$, $h_{nj}(\theta)$ converges to $h_j(\theta)$ and $\epsilon_{nj}(\theta)$ converges to 0 with probability one as $n \to \infty$ uniformly on $B$.
\end{assumption}

\begin{assumption} \label{assum:constraint-qualification}
For any $\theta \in \Theta$, let $\mathcal{A}(\theta) = \{j \colon h_j(\theta) = 0 \}$ be the indices of active constraints, which could be empty. We assume that the gradient vectors $\nabla h_j(\vartheta)$, $j \in \mathcal{A}(\vartheta)$, are linearly independent.
\end{assumption}

\section{Extension of \texorpdfstring{\textcite{Aronow2013}}{Aronow and Lee (2013)}} \label{sec:al-extension}
\subsection{Extension to ratio estimators} \label{sec:al-ratio-estimators}
Another sensitivity analysis that can be made more informative through the addition of auxiliary constraints is one proposed in \textcite{Aronow2013}. This non-parametric sensitivity analysis computes bounds on an inverse probability weighted sample mean under the assumption that each weight is bounded between two known constants. To start, we provide a slight generalization of this sensitivity analysis to ratio estimators. Let the estimator be given by
\begin{equation}
\beta_n = \displaystyle \left. \biggl(\sum_{i=1}^n f(T_i) /e(W_i) \biggr) \middle/ \biggr(\sum_{i=1}^n g(T_i) / e(W_i) \biggr) \right.
\label{eqn:ratio-estimator}
\end{equation}
where $T_i \in \mathcal{T} \subseteq \mathbb{R}^M$ and $f, g \colon \mathbb{R}^M \to \mathbb{R}$. We assume that each $e(W_i)$ is unknown (and some $W_i$ may be unmeasured) and lies between two user-specified constants $1 \leq a \leq e(W_i) \leq b < \infty$.

To apply our theoretical results in \Cref{sec:ci-for-constrained-pi}, we need a set $\Theta$ of fixed dimension. A simple assumption is that $\mathcal{T}$ is discrete and finite, $\mathcal{T} = \{t_k \colon k = 1, 2, \ldots, K\}$. Under this assumption, we can define $\beta_n(\theta)$ equal to
\[\begin{array}{ll}
& \beta_n(\theta) = \displaystyle \left. \biggl(\sum_{k=1}^K \theta_k f(t_k) p_n(t_k) \biggr) \middle/ \biggr(\sum_{k=1}^K \theta_k g(t_k) p_n(t_k) \biggr) \right. \\[1em]
& \beta(\theta) = \displaystyle \left. \biggl(\sum_{k=1}^K \theta_k f(t_k) p(t_k) \biggr) \middle/ \biggr(\sum_{k=1}^K \theta_k g(t_k) p(t_k) \biggr) \right.
\end{array}\]
where $\theta_k = E[1/e(W_i) \mid T_i = t_k, S_i = 1]$ and $p_n(\cdot), p(\cdot)$ are (respectively) the sample and population probability measures. This results in $\Theta = [1/b, 1/a]^K$. From here, the infimum takes the usual form,
\[
\nu_n = \inf \{ \beta_n(\theta) \colon \theta \in \Theta \}, \quad \nu = \inf \{ \beta(\theta) \colon \theta \in \Theta \}.
\]

It remains to identify assumptions such that the conditions in \Cref{sec:ci-for-constrained-pi} are satisfied.

\begin{assumption} \label{assum:positive-denominator}
For all $\theta \in \Theta$, the denominators of $\beta(\theta)$ and $\beta_n(\theta)$ are non-zero with probability one.
\end{assumption}

\begin{assumption} \label{assum:uniqueness-regularity}
For all $t_k \in \mathcal{T}$, $f(t_k)/g(t_k) \, \neq \, \nu$.
\end{assumption}

\Cref{assum:positive-denominator} simply ensures that $\beta_n(\theta)$ and $\beta(\theta)$ are well-defined over $\Theta$. \Cref{assum:uniqueness-regularity} is more subtle but will be needed to ensure a unique solution. If this assumption violated for some $k$, then the infimum is identical for all values of $\theta_k$, meaning that the solution is not unique. An example of a function violating this condition is the following:
\begin{equation*}
\beta(\theta) = \displaystyle \frac{-\theta_1 - 7 \theta_2 - 10 \theta_3}{\theta_1 + \theta_2 + \theta_3}, \quad \Theta = [1,2]^3
\end{equation*}
\noindent In this example, $\vartheta = (1,1,2)$ and $\vartheta = (1,2,2)$ are both minimizers over $\Theta$. The minimum value is -7 but $f(t_2)/g(t_2) = -7$ as well, which violates the condition. These assumptions jointly imply that $\nu$ has a unique minimizer. We begin with a technical lemma.
\begin{lemma} \label{lemma:al-minimiser}
$\nu_n = Q_n(\vartheta_n)$ is a global minimum over $\Theta$ if and only if, for all $k = 1,...,K$, $q_k f(t_k) \leq \, \nu \, q_k \, g(t_k)$, where $q_k = \vartheta_k - (1/a + 1/b - \vartheta_k)$.
\end{lemma} 

This leads to our main proposition.
\begin{proposition} \label{prop:al-uniqueness}
Under Assumptions \ref{assum:positive-denominator} and \ref{assum:uniqueness-regularity}, the set $\{ \theta \in \Theta \colon \beta(\theta) = \nu \}$ is a singleton and, for $n$ sufficiently large, the set $\{ \theta \in \Theta \colon \beta_n(\theta) = \nu_n \}$ is a singleton with probability one.
\end{proposition} 

In fact, \Cref{prop:al-uniqueness} provides an explicit form for the sample minimizer,
\begin{equation}
\vartheta_{nk} =
\begin{cases} 
      1/b & \text{ if }  f(t_k)/g(t_k) \geq \nu_n \\
      1/a & \text{ if }  f(t_k)/g(t_k) < \nu_n
   \end{cases},
\label{eqn:optweights}
\end{equation}
and the population minimizer takes a similar form but with $\nu_n$ replaced with $\nu$. Both Proposition 1 of \textcite{Aronow2013} and Section 4.4 of \textcite{Zhao2019} propose equivalent algorithms for computing the optimizing weights in the population means setting. \Cref{prop:al-uniqueness} shows that we can generalize this algorithm to ratio estimators. In short, we can order $f(t_k)/g(t_k)$ from smallest to largest and evaluate $\beta_n(\theta)$ by enumerating over the weight at which $1/b$ changes to $1/a$, which has computational complexity $O(n)$.

\Cref{prop:al-uniqueness} shows that \Cref{assum:uniqueness} is satisfied for the generalized \textcite{Aronow2013} estimator under relatively weak conditions. Furthermore, Assumptions \ref{assum:regularity}, \ref{assum:distribution-of-estimator} and \ref{assum:uniform-convergence} are satisfied under \Cref{assum:positive-denominator} and the assumption that $\mathcal{T}$ is finite. We therefore have that \Cref{prop:asymptotic-normality} is satisfied under these same relatively weak conditions.

\subsection{Auxiliary information constraints} \label{sec:al-auxiliary-constraints}
The constraints described in \Cref{ex:response-rate,ex:covariate-mean} can be applied to this estimator. Let $c_n =  Z_{\alpha_{1j}/2} \, n^{-1/2}$, then the response rate constraint can be formulated as
\begin{equation}
h_{nj}(\theta) = \bigl(1 + c_n^2 \bigr) \biggl\lbrace\sum_{k=1}^K (\theta_k - 1/r) p_n(t_k)\biggr\rbrace^2 - c_n^2 \sum_{k=1}^K (\theta_k - 1/r)^2 p_n(t_k) \leq 0.
\end{equation}
Suppose $w_k \in \mathbb{R}$ is an element of $t_k$, then the covariate mean constraint can be formulated as 
\begin{equation}
h_{nj}(\theta) = \bigl(1 + c_n^2 \bigr) \biggl\lbrace\sum_{k=1}^K \theta_k (w_k - \bar{w}) p_n(t_k)\biggr\rbrace^2 - c_n^2 \sum_{k=1}^K \theta_k^2 (w_k - \bar{w})^2 p_n(t_k) \leq 0.
\end{equation}
Both of these constraints are quadratic in $\theta$ and can therefore be solved by existing algorithms for quadratically-constrained linear programs. \Cref{ex:direction} cannot be extended to this setting because it is tied to a parametric model. \Cref{ex:negative-control} can be extended to this setting in principle, but the resulting optimization problem is intractable.

Uniqueness of $\vartheta$ over $\Theta$ is needed to invoke \Cref{thm:coverage}. The population minimization problem over $\Theta$ is a linearly-constrained linear fractional programming problem. For example, the population response rate constraint is
\[
h_j(\theta) = \sum_{k=1}^K (\theta_k - 1/r) p(t_k),
\]
which is linear in $\theta$. Since the level sets of $\beta(\theta)$ are also linear, to establish uniqueness of $\vartheta$ it suffices to assume (in addition to Assumptions \ref{assum:positive-denominator} and \ref{assum:uniqueness-regularity}) that the coefficient vectors of $h_j(\theta)$ and the level sets of $\beta(\theta)$ over $\Theta$ are not parallel. 

\section{Technical details} \label{sec:appendix:proofs}

\subsubsection*{Proof of \Cref{lemma:al-minimiser}}
\begin{proof}
Consider the population problem. We will prove the `if' statement since the `only if' statement follows from some simple algebra. We will begin by noting that, since $\nu$ is the solution to a linear fractional programming problem and since $\Theta$ is a compact, convex polyhedron, the maximizing weight vector $\vartheta$ will lie at a vertex. In other words, $\vartheta \in \lbrace 1/b,\,1/a \rbrace^K$. Take an arbitrary weight vector $\theta \in \lbrace 1/b,\,1/a \rbrace^K$. Suppose there are $1 < m \leq K$ elements of $\theta$ which differ from $\vartheta$. Without loss of generality, suppose these are the first $m$ elements. Then we can write

\begin{align*}
\beta(w) &= \displaystyle \frac{\sum_{k=1}^K \theta_k f(t_k) p(t_k) }{\sum_{k=1}^K \theta_k g(t_k) p(t_k)} \\[1em]
&= \displaystyle \frac{\sum_{k=1}^K \vartheta_k f(t_k) p(t_k)  -  \sum_{k=1}^m q_k f(t_k) p(t_k) }{\sum_{k=1}^K \vartheta_j g(t_k) p(t_k) - \sum_{k=1}^m q_k g(t_k) p(t_k)} \\[1em]
&  \geq \displaystyle \frac{ \nu \, \sum_{k=1}^K \vartheta_k g(t_k) p(t_k) - \nu \, \sum_{k=1}^m q_k  g(t_k)  p(t_k) }{\sum_{k=1}^K \vartheta_k g(t_k)  p(t_k)  - \sum_{k=1}^m q_k g(t_k) p(t_k)} \\[1em]
&= \displaystyle \nu \, \frac{ \sum_{k=1}^K \vartheta_k g(t_k) p(t_k) - \sum_{k=1}^m q_k g(t_k) p(t_k) }{\sum_{k=1}^K \vartheta_k g(t_k) p(t_k) - \sum_{k=1}^m q_k g(t_k) p(t_k)} \\[1em]
& = \nu
\end{align*}
The same holds with probability one for the sample problem by replacing $p(t_k)$ with $p_n(t_k)$.
\end{proof}

\subsubsection*{Proof of \Cref{prop:al-uniqueness}}
\begin{proof}
Consider the population problem. Suppose that there are two global minima, $\nu_1 = \beta(\vartheta_1)$ and $\nu_2 = \beta(\vartheta_2)$, such that $\nu_1 = \nu_2 = \nu$ and $\vartheta_1 \neq \vartheta_2$. Since $\nu_1$ and $\nu_2$ are both global minima then, by \Cref{lemma:al-minimiser}, for all $k = 1,\ldots,K$,
\begin{equation}
\begin{array}{l@{}l}
&{} q_k \, f(t_k) \leq \nu_1 \, q_k \, g(t_k)\\[0.5em]
&{} q_k \, f(t_k) \leq \nu_2 \, q_k \, g(t_k)
\end{array}
\label{eqn:al-inequality}
\end{equation}
Without loss of generality, we assume that $\vartheta_1$ and  $\vartheta_2$ differ by the first $m$ elements. Then,
\begin{equation*}
\begin{array}{l@{}l}
\nu = \nu_1 &{}= \displaystyle \frac{\sum_{k=1}^K \vartheta_{1j} f(t_k) p(t_k) }{\sum_{k=1}^K \vartheta_{1k} g(t_k) p(t_k)} \\[1.5em]
&{}= \displaystyle \frac{\sum_{k=1}^K \vartheta_{2k} f(t_k) p(t_k) }{\sum_{k=1}^K \vartheta_{2k} g(t_k) p(t_k)} \\[1.5em]
&{}= \displaystyle \frac{\sum_{k=1}^K \vartheta_{1k} f(t_k) p(t_k) - \sum_{k=1}^K q_k f(t_k) p(t_k)}{\sum_{k=1}^K \vartheta_{1k} g(t_k) p(t_k) - \sum_{k=1}^m q_k g(t_k) p(t_k)}
\end{array}
\end{equation*}
where $q_k = \vartheta_{1k} - (1/a + 1/b - \vartheta_{1k})$. Rearranging, we obtain,
\begin{equation*}
\displaystyle \sum_{k=1}^m q_k \, f(t_k) \, p(t_k) \, = \, \nu \, \sum_{k=1}^m q_k \, g(t_k) \, p(t_k)
\end{equation*}
However, \eqref{eqn:al-inequality} implies that this equality will only hold if, for all $k = 1,...,m$, $f(t_k)/g(t_k) = \nu$, which cannot be true by Assumption \ref{assum:uniqueness}. Therefore, by contradiction, the set $\{\theta \in \Theta \colon \beta(\theta) = \nu \}$ must be a singleton. The same holds with probability one for the sample problem by replacing $p(t_k)$ with $p_n(t_k)$.
\end{proof}

\subsubsection*{Proof of \Cref{prop:variance-consistency}}
\begin{proof}
\begin{equation*}
\begin{array}{l@{}l}
\vert \sigma_n^2(\vartheta_n) - \sigma^2(\vartheta)  \vert &{}= \vert  \sigma_n^2(\vartheta_n) - \sigma^2(\vartheta_n) + \sigma^2(\vartheta_n) - \sigma^2(\vartheta) \vert \\[1em]

&{}\leq \vert \sigma_n^2(\vartheta_n) - \sigma^2(\vartheta_n) \vert \, + \, \vert \sigma^2(\vartheta_n) - \sigma^2(\vartheta) \vert \\[1em]

&{}\leq \displaystyle \sup_{\theta \in \Theta} \vert \sigma_n^2(\theta) - \sigma^2(\theta) \vert \, + \, \vert \sigma^2(\vartheta_n) - \sigma^2(\vartheta) \vert \\[1em]

&{}\to 0 ~ \text{with probability one},
\end{array}
\end{equation*}
where the last inequality holds by the uniform strong consistency and the second term on the last line goes to zero with probability one since $\vartheta_n \, \to \, \vartheta$ with probability one by \Cref{prop:consistency} and $\sigma^2(\theta) \in C(S)$.
\end{proof}

Before proving \Cref{thm:coverage}, we begin with some notation and preliminary lemmas. Denote the confidence bound for a particular $\theta$ and $\alpha$ as
\[
C_n(\theta, \alpha) = Q_n(\theta) - Z_{\alpha} \, \sigma_n(\theta) \, n^{-1/2} 
\]
and the sample minimum over $\Theta_n^r$ at $\alpha_2$ as
\[
\zeta_n^r \in \argmin \{ C_n(\theta, \alpha_2) \colon \theta \in \Theta_n^r \}.
\]
Recall that $\vartheta_n^r \in \argmin \{ Q_n(\theta) \colon \theta \in \Theta^r \}$ and $\vartheta = \argmin \{ Q(\theta) \colon \theta \in \Theta \}$, which is assumed to be unique. These quantities can be ordered deterministically as
\[
C_n(\zeta_n^r, \alpha) \, \leq \, C_n(\vartheta_n^r, \alpha) \, \leq \, Q_n(\vartheta_n^r) \, \leq \, Q_n(\zeta_n^r).
\]
The first lemma provides a lower bound for the coverage probability of $C_n(\vartheta_n^r, \alpha_2)$.

\begin{lemma} \label{lemma:main-theorem-1}
Let $\delta_n = Z_{\alpha_2} \, \epsilon \, n^{-1/2}$ be a deterministic sequence where $\epsilon > 0$ is any positive constant, then
\begin{align*}
\pr\biggl( C_n(\vartheta_n^r, \alpha_2) \, \leq \, \nu \biggr) \, \geq \, & \pr\biggl( C_n(\zeta_n^r, \alpha_2) \, \leq \, \nu - \delta_n \biggr) \, + \, \pr \biggl( \bigl\vert \sigma_n(\zeta_n^r) \, - \, \sigma_n(\vartheta_n^r) \bigr\vert \, \leq \, \epsilon \biggr) \, - \, 1. 
\end{align*}
\end{lemma}

\begin{proof}
\begin{align*}
&\pr\biggl( C_n(\vartheta_n^r, \alpha_2) \, \leq \, \nu \biggr) \\[1em] = \, &\pr\biggl( Q_n(\vartheta_n^r) \, - \, Z_{\alpha_2} \, \sigma_n(\vartheta_n^r) \, n^{-1/2} \, \leq \, \nu \biggr) \\[1em]
= \, &\pr\biggl( Q_n(\zeta_n^r) \, - \, Z_{\alpha_2} \sigma_n(\zeta_n^r) n^{-1/2} \, + \, \bigl\lbrace (Q_n(\vartheta_n^r) - Q_n(\zeta_n^r) \bigr\rbrace \, + \, Z_{\alpha_2} \bigl\lbrace \sigma_n(\zeta_n^r) \, - \, \sigma_n(\vartheta_n^r) \bigr\rbrace n^{-1/2}  \, \leq \, \nu \biggr) \\[1em]
\geq \, &\pr\biggl(Q_n(\zeta_n^r) \, - \, Z_{\alpha_2} \sigma_n(\zeta_n^r) n^{-1/2} \, + \, Z_{\alpha_2} \bigl\lbrace \sigma_n(\zeta_n^r) \, - \, \sigma_n(\vartheta_n^r) \bigr\rbrace n^{-1/2}  \, \leq \, \nu \biggr) \\[1em]
\geq \, &\pr\biggl( Q_n(\zeta_n^r) \, - \, Z_{\alpha_2} \sigma_n(\zeta_n^r) n^{-1/2} \, + \, \delta_n  \, \leq \, \nu, \, Z_{\alpha_2} \bigl\vert \sigma_n(\zeta_n^r) \, - \, \sigma_n(\vartheta_n^r) \bigr\vert n^{-1/2} \, \leq \, \delta_n \biggr) \\[1em]
\geq \, &\pr\biggl( C_n(\zeta_n^r, \alpha_2) \, \leq \, \nu - \delta_n \biggr) \, + \, \pr \biggl( Z_{\alpha_2} \bigl\vert \sigma_n(\zeta_n^r) \, - \, \sigma_n(\vartheta_n^r) \bigr\vert n^{-1/2} \, \leq \, \delta_n \biggr) \, - \, 1 \\[1em]
= \, &\pr\biggl( C_n(\zeta_n^r, \alpha_2) \, \leq \, \nu - \delta_n \biggr) \, + \, \pr \biggl( \bigl\vert \sigma_n(\zeta_n^r) \, - \, \sigma_n(\vartheta_n^r) \bigr\vert \, \leq \, \epsilon \biggr) \, - \, 1.
\end{align*}
\end{proof}

\begin{lemma} \label{lemma:consistency}
Suppose a sequence of functions $\tilde{Q}_n \colon \mathbb{R}^p \to \mathbb{R}$ is in $C(B)$ and converges to $Q(\theta)$ with probability one as $n \to \infty$ uniformly on $B$. Furthermore, let $\tilde{\nu}_n = \inf \{\tilde{Q}_n(\theta) \colon \Theta_n^r \}$ and $\tilde{\vartheta}_n \in \{\theta \colon \tilde{Q}_n(\theta) = \tilde{\nu}_n \}$. Then, under Assumptions \ref{assum:uniqueness}, \ref{assum:regularity} and \ref{assum:compact-subset} - \ref{assum:constraint-qualification}, $\tilde{\nu}_n \to \nu$ and $\tilde{\vartheta}_n \to \vartheta$ with probability one as $n \to \infty$.
\end{lemma}

\begin{proof}
The proof of this lemma combines Theorem 5.3 and Theorem 5.5 in \textcite{Shapiro2009}. Theorem 5.3 establishes consistency of optimal values and solutions when $\Theta$ is known. Theorem 5.5 generalizes this result to an estimated constraint set, in our case $\Theta_n^r$.

In particular, Theorem 5.5 requires that the following two conditions are satisfied:
\begin{enumerate}[label=\alph*)]
    \item If $\theta_n \in \Theta_n^r$ and $\theta_n$ converges with probability one to a point $\theta^*$, then $\theta^* \in \Theta$.
    \item There exists a sequence $\theta_n \in \Theta_n^r$ such that $\theta_n$ converges to $\vartheta$ with probability one.
\end{enumerate}

We begin with the proof of condition (a). Suppose $\theta^* \not\in \Theta$. Then there exists some $j = 1, \ldots, J$ such that $h_j(\theta^*) \geq \delta$, where $\delta > 0$ is some constant. By the triangle inequality,
\[
| h_{jn}(\theta_n) - h_j(\theta^*) | \leq | h_{jn}(\theta_n) - h_{jn}(\theta^*) |  + | h_{jn}(\theta^*) - h_j(\theta^*) |.
\]
The first term converges to zero with probability one because $\theta_n$ converges to $\theta^*$ and $h_{jn}(\theta) \in C(B)$, both with probability one. The second term also converges to zero with probability one by \Cref{assum:constraint-convergence} because $h_{jn}(\theta) \to h_j(\theta)$ uniformly on $B$ with probability one. 
This means that for all $\epsilon > 0$ there exists an $n \geq n_{\epsilon}$ such that
\[
| h_{jn}(\theta_n) - h_j(\theta^*) | < \epsilon.
\]
However, since $h_j(\theta^*) \geq \delta$ and $\theta_n \in \Theta_n^r$, such that $h_{jn}(\theta_n) \leq \epsilon_{jn}(\theta_n)$, this is equivalent to
\[
h_j(\theta^*) - h_{jn}(\theta_n) < \epsilon
\]
whenever $\delta > \epsilon_{jn}(\theta_n)$. Without loss of generality, we can set $\epsilon = \delta - \epsilon_{jn}(\theta) > 0$. From here, we can rearrange,
\[
h_{jn}(\theta_n) > h_j(\theta^*) - \epsilon \geq \delta - \epsilon \geq \epsilon_{jn}(\theta_n),
\]
which means that $\theta_n \not\in \Theta_n^r$, which is a contradiction. Therefore, it must be that $\theta^* \in \Theta$.

Condition (b) follows from the constraint qualification imposed in \Cref{assum:constraint-qualification} and the discussion in \textcite[p. 161-162]{Shapiro2009}.
\end{proof}

\subsubsection*{Proof of Theorem \ref{thm:coverage}}
\begin{proof}

We begin by invoking \Cref{lemma:main-theorem-1}, which states that
\begin{align*}
\pr\biggl( C_n(\vartheta_n^r, \alpha_2) \, \leq \, \nu \biggr) \, \geq \, & \pr\biggl( C_n(\zeta_n^r, \alpha_2) \, \leq \, \nu - \delta_n \biggr) \, + \, \pr \biggl( \bigl\vert \sigma_n(\zeta_n^r) \, - \, \sigma_n(\vartheta_n^r) \bigr\vert \, \leq \, \epsilon \biggr) \, - \, 1,
\end{align*}
where $\delta_n = Z_{\alpha_2} \, \epsilon \, n^{-1/2}$ and $\epsilon > 0$ is any positive constant. We claim that 
\[
\lim_{n \to \infty} \pr \biggl( \bigl\vert \sigma_n(\zeta_n^r) \, - \, \sigma_n(\vartheta_n^r) \bigr\vert \, \leq \, \epsilon \biggr) = 1.
\]

This follows from \Cref{prop:variance-consistency} and \Cref{assum:variance-closeness}. A sufficient condition for satisfying this assumption is that $\vartheta_n^r \to \vartheta$ and $\zeta_n^r \to \vartheta$ with probability one. This follows from \Cref{lemma:consistency} since $Q_n(\theta)$ and $C_n(\theta, \alpha)$ both converge to $Q(\theta)$ with probability one uniformly on $B$ by \Cref{assum:uniform-convergence}. Therefore, we have that 
\begin{align*}
& \displaystyle \lim_{n \to \infty} \, \pr\biggl( C_n(\vartheta_n^r, \alpha_2) \, \leq \, \nu \biggr) \\[1em]
\geq & \displaystyle \lim_{n \to \infty} \, \pr\biggl( C_n(\zeta_n^r, \alpha_2) \, \leq \, \nu - \delta_n \biggr) \\[1em]
\geq & \displaystyle \lim_{n \to \infty} \, \pr\biggl( C_n(\zeta_n^r, \alpha_2) \, \leq \, \nu - \delta_n, ~ \Theta \, \subseteq \, \Theta_n^r \biggr) \\[1em]
\geq & \displaystyle \lim_{n \to \infty} \, \pr\biggl( C_n(\vartheta_n, \alpha_2) \, \leq \, \nu - \delta_n, ~ \Theta \, \subseteq \, \Theta_n^r \biggr) \\[1em]
= & \displaystyle \lim_{n \to \infty} \, \biggl\lbrace \pr\biggl( Q_n(\vartheta_n) - Z_{\alpha_2} \sigma_n(\vartheta_n) n^{-1/2} \, \leq \, \nu - \delta_n \biggr) - \displaystyle \pr\biggl( C_n(\vartheta_n, \alpha_2) \, \leq \, \nu - \delta_n, ~ \Theta \, \not\subseteq \, \Theta_n^r \biggr) \biggr\rbrace \\[1.5em]
\geq & \displaystyle \lim_{n \to \infty} \, \biggl\lbrace \pr\biggl( n^{1/2} \, \frac{\bigl(Q_n(\vartheta_n) \, - \, \nu \bigr)}{\sigma(\vartheta)} \, \frac{\sigma(\vartheta)}{\sigma_n(\vartheta_n)} \, \leq \, Z_{\alpha_2} (1 - \epsilon ) \biggr) \, - \, \pr\biggl( \Theta \, \not\subseteq \, \Theta_n^r \biggr) \biggr\rbrace \\[1em]
\geq & \, \Phi(Z_{\alpha_2} (1 - \epsilon)) \, - \, \alpha_1
\end{align*}
where the last inequality follows by Slutsky's theorem, \Cref{prop:variance-consistency} and \Cref{prop:asymptotic-normality}. Since $\epsilon > 0$ is an arbitrarily small constant, this lower bound can be set arbitrarily close to $1 - \alpha_2 - \alpha_1$.
\end{proof}

\end{document}